\pdfoutput=1
\documentclass[a4paper]{jpconf}
\usepackage{graphicx}
\usepackage{natbib}
\newcommand{\bm}[1]{\mbox{\protect\boldmath$#1$}}

\begin{document}

\title{Frame-dragging effects on magnetic fields near a rotating black hole}

\author{
V. Karas, O. Kop\'a\v{c}ek and D. Kunneriath
}

\address{Astronomical Institute, Academy of Sciences, Bo\v{c}n\'{\i} II 1401, CZ-14131 Prague, Czech Republic}

\ead{vladimir.karas@cuni.cz}

\begin{abstract}
We discuss the role of general relativity frame dragging acting on 
magnetic field lines near a rotating (Kerr) black hole. Near ergosphere the 
magnetic structure becomes strongly influenced and magnetic null 
points can develop. We consider aligned magnetic fields as well as 
fields inclined with respect to the rotation axis, and the two cases are
shown to behave in profoundly different ways. Further, we construct
surfaces of equal values of local electric and magnetic intensities,
which have not yet been discussed in the full generality of a boosted
rotating black hole.
\end{abstract}

\section{Introduction}
Various processes can lead to the magnetic topology exhibiting 
null points, in particular, these can emerge by complex motions
of the plasma. In this respect the black hole rotation 
brings a new situation, as magnetic field lines become twisted in a 
highly curved spacetime of a rotating black hole, approximately
where the ergosphere forms. Here, we
examine the resulting structure of the magnetic field, namely, the 
emergence of critical points in a local frame of a physical observer,
resembling the occurrence of X-points. 

Previously \citep{kk09} we considered a special case
of a uniform magnetic field in perpendicular orientation with respect 
to the black hole spin axis, and we demonstrated that magnetic 
null points can indeed form near a rotating (Kerr) black hole. 
Here, the embedded magnetic field is allowed a
general orientation, i.e., it can be inclined in an arbitrary angle with respect
to the rotation axis. The axial symmetry is broken between the gravitational 
and electromagnetic fields, and this has to result in a truly
three-dimensional structure of magnetic field lines along with 
gravito-magnetically induced electric field. The adopted approach
allows us to identify the gravitational effects operating in 
a magnetically dominated system, where a super-equipartion magnetic field
governs the motion of plasma. We further expand the discussion
in \citet{kk12} by constructing surfaces of equal values of local electric 
and magnetic intensities, even for the case of a general (oblique) 
orientation of the background magnetic field.

\section{Aligned and oblique magnetic fields near black hole horizon}
Astrophysical black holes do not support their own intrinsic 
magnetic field; this has to be generated by external currents and 
brought down to horizon by accretion. A black hole can also 
enter a pre-existing magnetic flux tube, and then one asks if the 
process of magnetic reconnection is influenced by strong gravitational 
field near horizon. And does the black hole rotation play a significant 
role in the scenario of this kind?

As we wish to discuss magnetic fields inclined with respect to
the spin axis, and we also want to include the fast 
translatory motion, the following picture appears to be appropriate:
Kerr black hole traversing a ``magnetic filament'', described as an
extended 
(largely one-dimensional) flux tube. Such an idea can be motivated
observationally, by highly ordered and elongated arcs (of 
about $100\mu$G) that are seen in Galactic Center, 
within a few parsecs from Sagittarius 
A* compact radio source \citep[Sgr~A*;][]{ym84}. They are thought to represent 
large-scale magnetic flux tubes that are illuminated by
synchrotron emission from relativistic electrons \citep{ln04,m06}.

Given a limited resolution that can be achieved with current
imaging techniques, the magnetic filaments cannot be traced down to the 
characteristic size of the black hole.\footnote{Gravitational radius
$r_{\rm{g}}=GM/c^2\approx1.48\times10^{13}\,M_8$~cm, where the central
black hole mass is expressed in terms of $M_8{\equiv}M/(10^8M_{\odot})$.
The velocity of the Keplerian orbital motion of a particle 
is then $v_{_{\rm{}K}}\approx2.1\times10^{10}
(r/r_{\rm{}g})^{-1/2}{\rm{cm\;s}}^{-1}$. The corresponding orbital
period is $T_{_{\rm{}K}}\approx3.1\times10^3(r/r_{\rm{}g})^{3/2}M_8$~s.
Hereafter, we use a dimensionless form of geometrized units,
where all quantities are scaled with the black hole mass;
$M$ does not appear in the equations explicitly.} Therefore,
the actual mapping of the magnetic structures near black holes 
is not directly possible. Nonetheless, their role in accelerating the 
particles is suggestive, especially because they could help us 
to understand the origin of particle acceleration and the resulting 
signatures in the electromagnetic signal.

Naturally, in astrophysically 
realistic solutions the role of non-ideal plasma will need be included. Currently, 
neither of the frequently discussed limits (i.e., vacuum vs.\ force-free
approximations) are able to account for both the plasma currents 
as well as the accelerating electric fields. This task will require
resistive MHD, which is, however, beyond the scope of this brief 
paper.

\subsection{The effect of black hole rotation and translatory motion}
Starting from  \citet{kl75} and \citet{bd76}, the organised 
electromagnetic test fields surrounding 
black holes have been discussed and their astrophysical consequences
considered in various papers. The case of oblique 
geometry, however, has been explored only partially \citep{bj80}. 
This is caused especially by the fact that the off-equatorial fields
are lacking any symmetry, and so are more difficult to visualise. 
Also, qualitatively new effects on the field 
structure were not expected. 

Nevertheless, in \citet{kk09} we were 
able to demonstrate that the asymptotically perpendicular field 
lines develop a magnetic neutral point in the equatorial plane. 
This is interesting because such structures of the magnetic field
are relevant for processes of electromagnetic acceleration. 
The magnetic null point emerges in a local 
physical frame, and could trigger reconnection. 
However, the asymptotically perpendicular field is a rather 
exceptional case. Therefore, here we investigate near-horizon 
magnetic structures in more detail, also for a general 
orientation of the magnetic field (see also \citeauthor{kk12}
\citeyear{kk12}).

\begin{figure}[tbh!]
\centering
\includegraphics[width=0.49\textwidth]{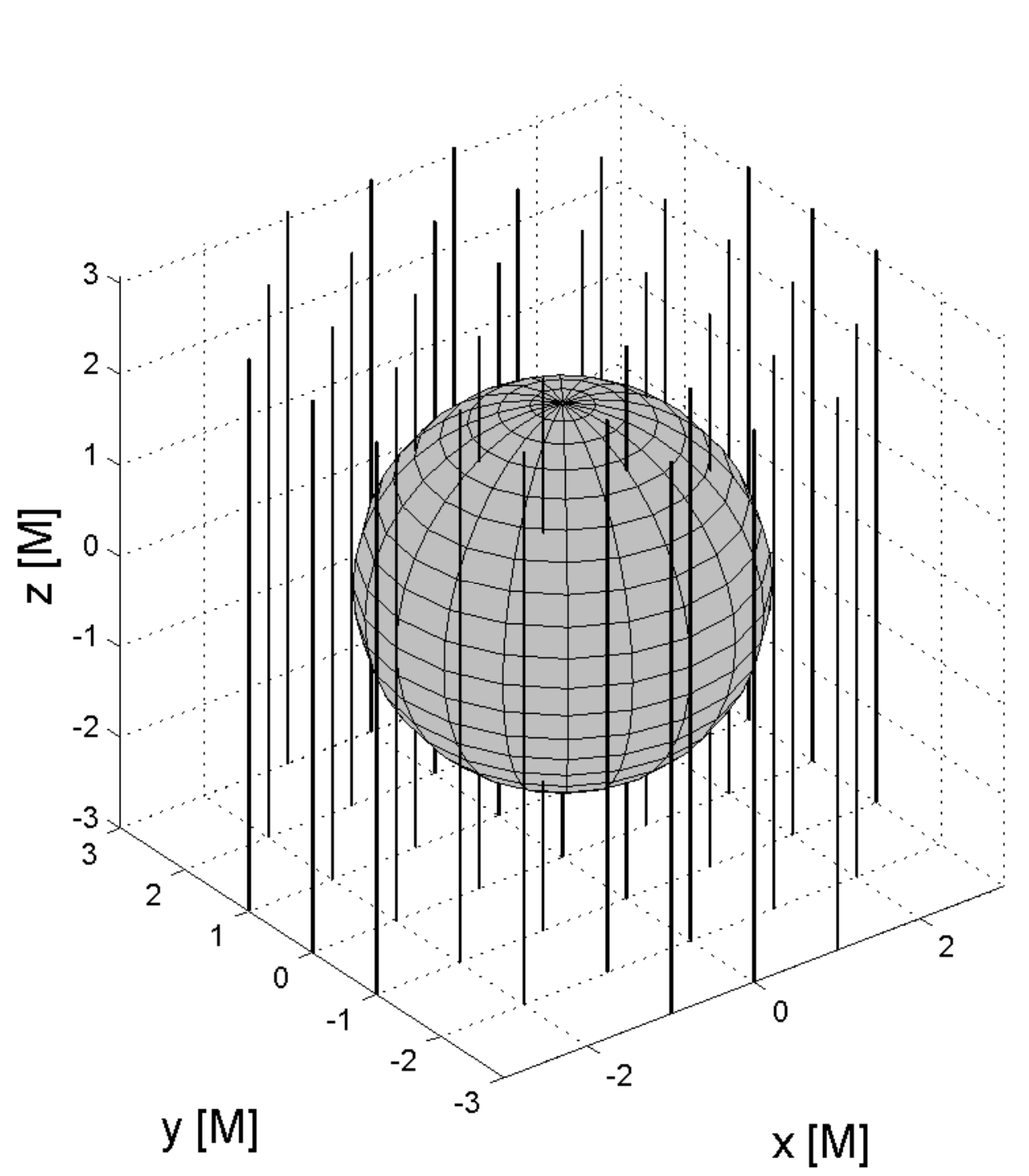}
\hfill
\includegraphics[width=0.49\textwidth]{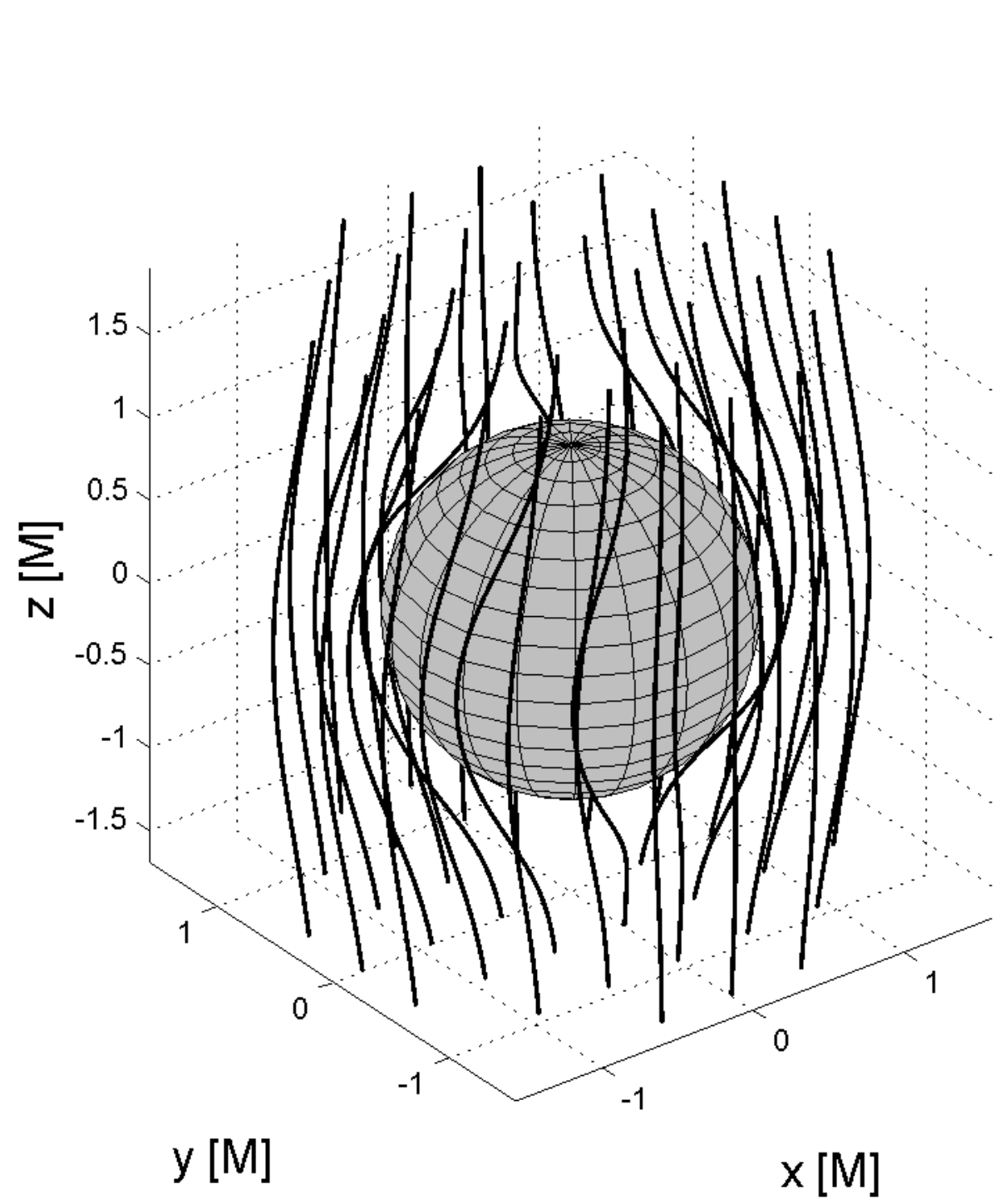}
\caption{Magnetic lines of a uniform field near a non-rotating black hole. 
Left: The field lines appear to
be perfectly homogeneous (in these suitably defined coordinates),
plotted with respect to frame of free-falling observers.
Right: magnetic field aligned with the rotation axis of an maximally 
rotating ($a=M$) Kerr black hole, when they are expelled out of
the horizon. 
\citep[see][for more details]{k11}.}
\label{fig1}
\end{figure}

The main objective of this discussion is to track the location of
magnetic null points and to explore a complex three-dimensional
configuration inside the ergosphere. We consider the form of magnetic 
lines together with the induced electric lines for different 
values of the model parameters: the inclination angle of the asymptotic 
magnetic field $\theta_{\rm{}o}$ ($=\arctan(B_{\perp}/B_{\parallel}$), 
the black hole spin $a$ ($a^2\leq M$), and the boost velocity $\beta$ 
($\beta^2\equiv v_x^2+v_y^2+v_z^2<1$). We observe the layers of 
alternating magnetic orientation to occur also in the general case, i.e., 
when the black hole rotates and moves with respect to an oblique
magnetic field. However, the three-dimensional structure of the field
lines is very complicated as they become highly entangled around the 
null point. 

We specify the gravitational field by Kerr metric.
Our starting form of the 
electromagnetic field is a stationary solution of Maxwell's test-field 
electro-vacuum equations in the given curved spacetime.\footnote{These are 
astrophysically acceptable approximations which reflect the fact that 
black holes can only acquire a negligibly small electric charge, while 
cosmic electromagnetic fields are not strong enough to influence the 
spacetime metric significantly.} The electromagnetic four-potential $\bm{A}$ 
can be then written as superposition of two contributions: 
$\bm{A}=B_{\parallel}\,\bm{a}_{\parallel}+B_{\perp}\,\bm{a}_{\perp}$,
where $B_{\parallel}$ and $B_{\perp}$ define the magnitudes of the two
parts.

\begin{figure}[tbh!]
\centering
\hfill
\includegraphics[scale=0.59, clip,trim=30mm 5mm 65mm 5mm]{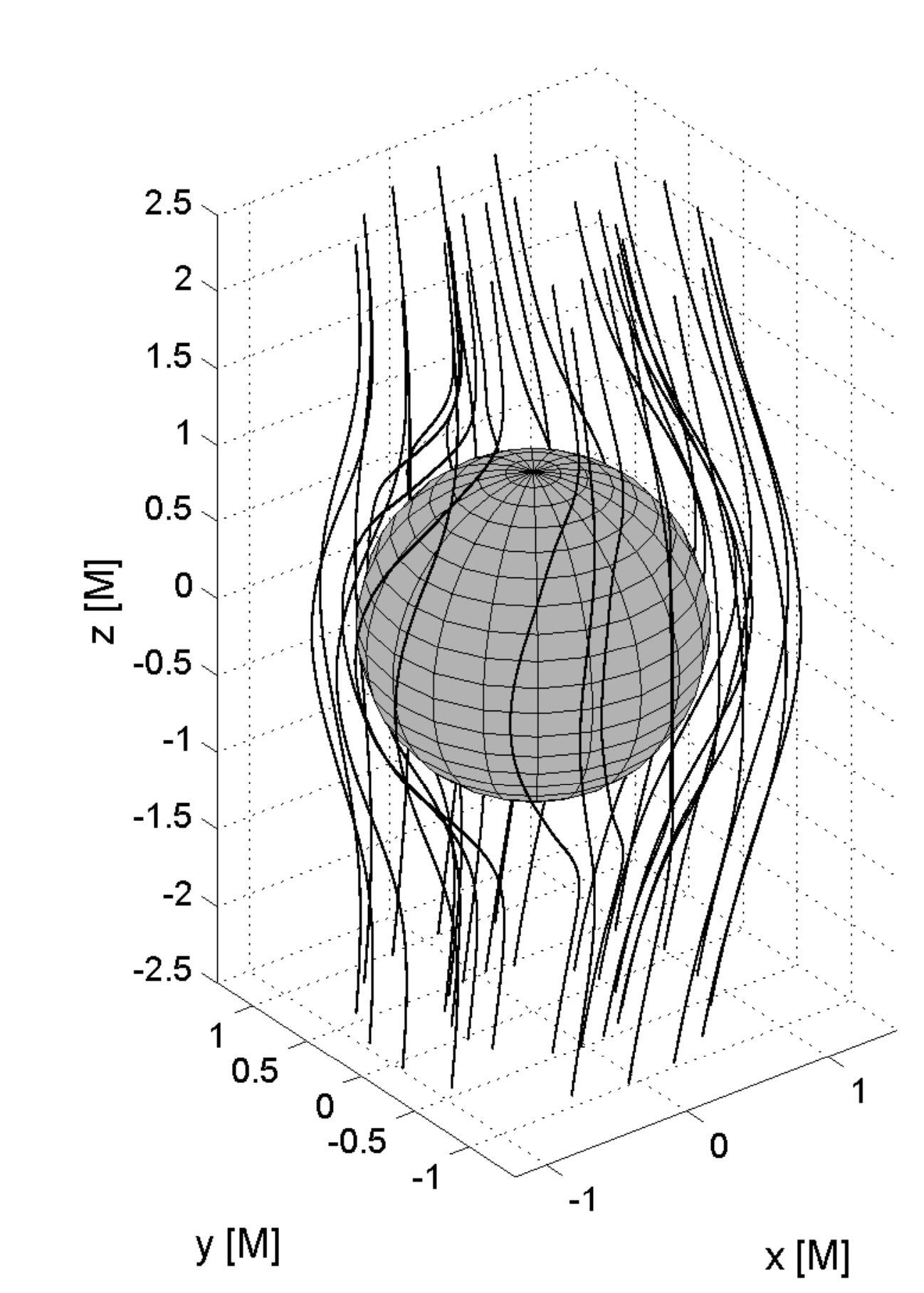}
\hfill
\includegraphics[scale=0.59, clip,trim=30mm 5mm 65mm 5mm]{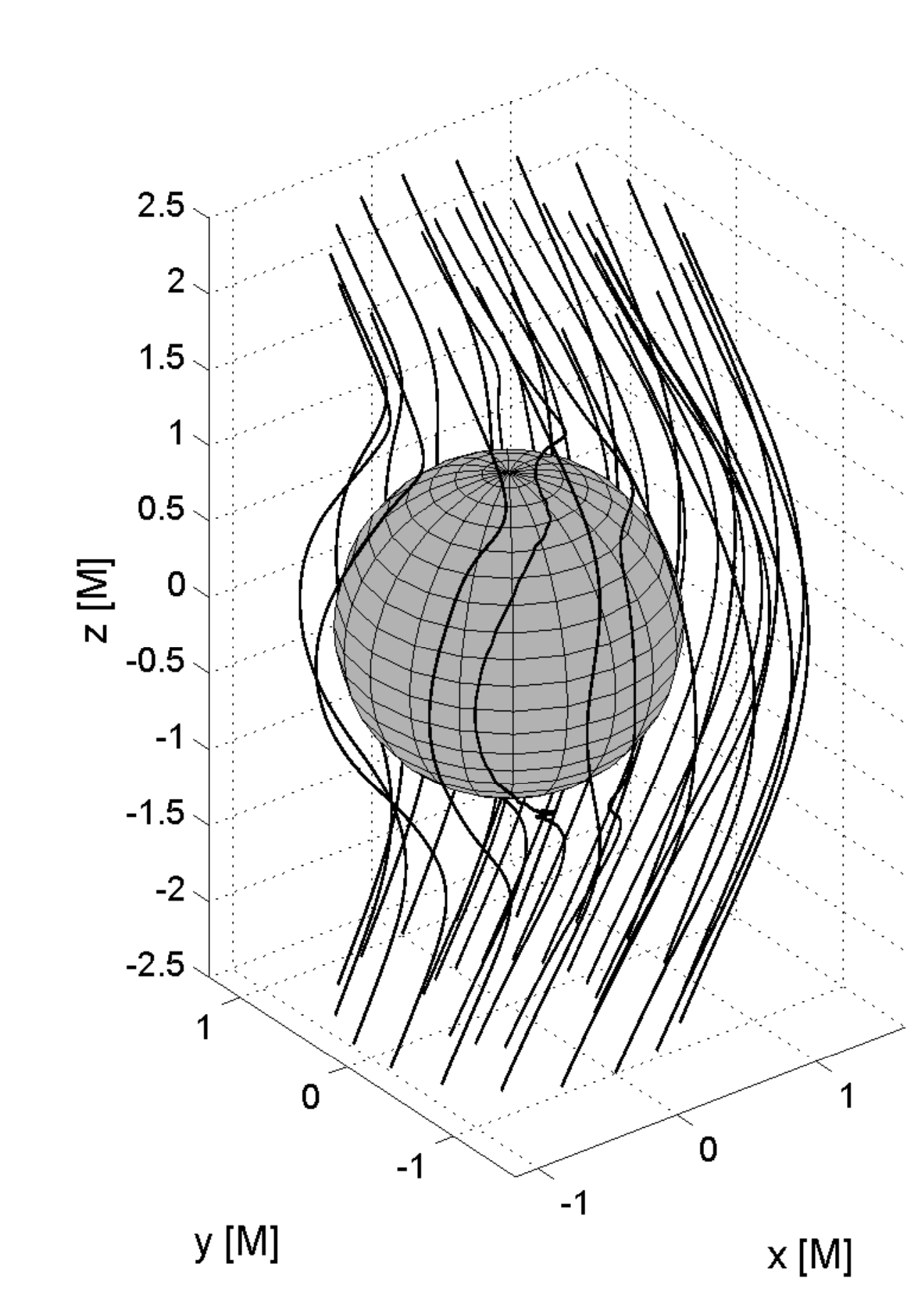}
\hfill~
\caption{The case of extreme spin, $a=M$, taking into account non-zero velocity 
of the translatory boost along $x$-direction: $v_x=0.1$ (left),  $v_x=0.3$ (right).}
\label{fig2}
\end{figure}

The aligned field has two non-vanishing components of the normalized 
electromagnetic four-potential,
\begin{eqnarray}
a_{t\parallel} &=& B_{\parallel}a\Big[r\Sigma^{-1}\left(1+\mu^2\right) -1\Big], \label{mf1}\\
a_{\phi\parallel} &=& B_{\parallel}\Big[{\textstyle\frac{1}{2}}\big(r^2+a^2\big)
 -a^2r\Sigma^{-1}\big(1+\mu^2\big)\Big] \sigma^2 \label{mf2},
\end{eqnarray}
where we use Boyer-Lindquist $t$, $r$, $\theta$, and $\phi$ 
dimension-less spheroidal coordinates ($\mu=\cos\theta$, $\sigma=\sin\theta$). 
Eqs.\ (\ref{mf1})--(\ref{mf2}) represent an asymptotically homogeneous 
magnetic field \citep{w74}. 

On the other hand, the perpendicular to axis component of the field is 
given by \citep{bj80}
\begin{eqnarray}
a_{t\perp} &=& B_{\perp}a\Sigma^{-1}\Psi\sigma\mu, \label{mf3} \\
a_{r\perp} &=& -B_{\perp}(r-1)\sigma\mu\sin\psi, \\
a_{\theta\perp} &=& -B_{\perp}\Big[\big(r\sigma^2+\mu^2\big) a\cos\psi 
 + \Big(r^2\mu^2+\big(a^2-r\big)(\mu^2-\sigma^2)\Big) \sin\psi\Big], \\
a_{\phi\perp} &=& -B_{\perp}\Big[\Delta\cos\psi+\big(r^2+a^2\big)
 \Sigma^{-1}\Psi\Big] \sigma\mu, \label{mf4}
\end{eqnarray}
with $\Sigma(r,\mu)$ and $\Delta(r)$ being the Kerr metric functions,
$\psi\equiv\phi+a\delta^{-1}\ln\left[\left(r-r_+\right)/\left(r-r_-\right)\right]$, 
$\Psi=r\cos\psi-a\sin\psi$, $\delta=r_+-r_-$, and $r_{\pm}=1\pm\sqrt{1-a^2}$.
Knowing the complete set of four-potential components the magnetic field structure
is fully determined: $F_{\mu\nu}\equiv A_{[\mu,\nu]}$.

Equations (\ref{mf3})--(\ref{mf4}) describe the field lacking the axial 
symmetry. Such a situation cannot be strictly stationary, however, 
the alignment time-scale is very long, and so we can neglect the 
associated energy losses here. The field line structure 
depends also on the motion of observers determining the field 
components. In order to construct and discuss our plots,
we employ a free-falling physical frame, evaluate the 
electromagnetic tensor with respect to the local frame, and use 
these components to draw the field lines.

\begin{figure}[tbh!]
\centering
\includegraphics[width=0.85\textwidth]{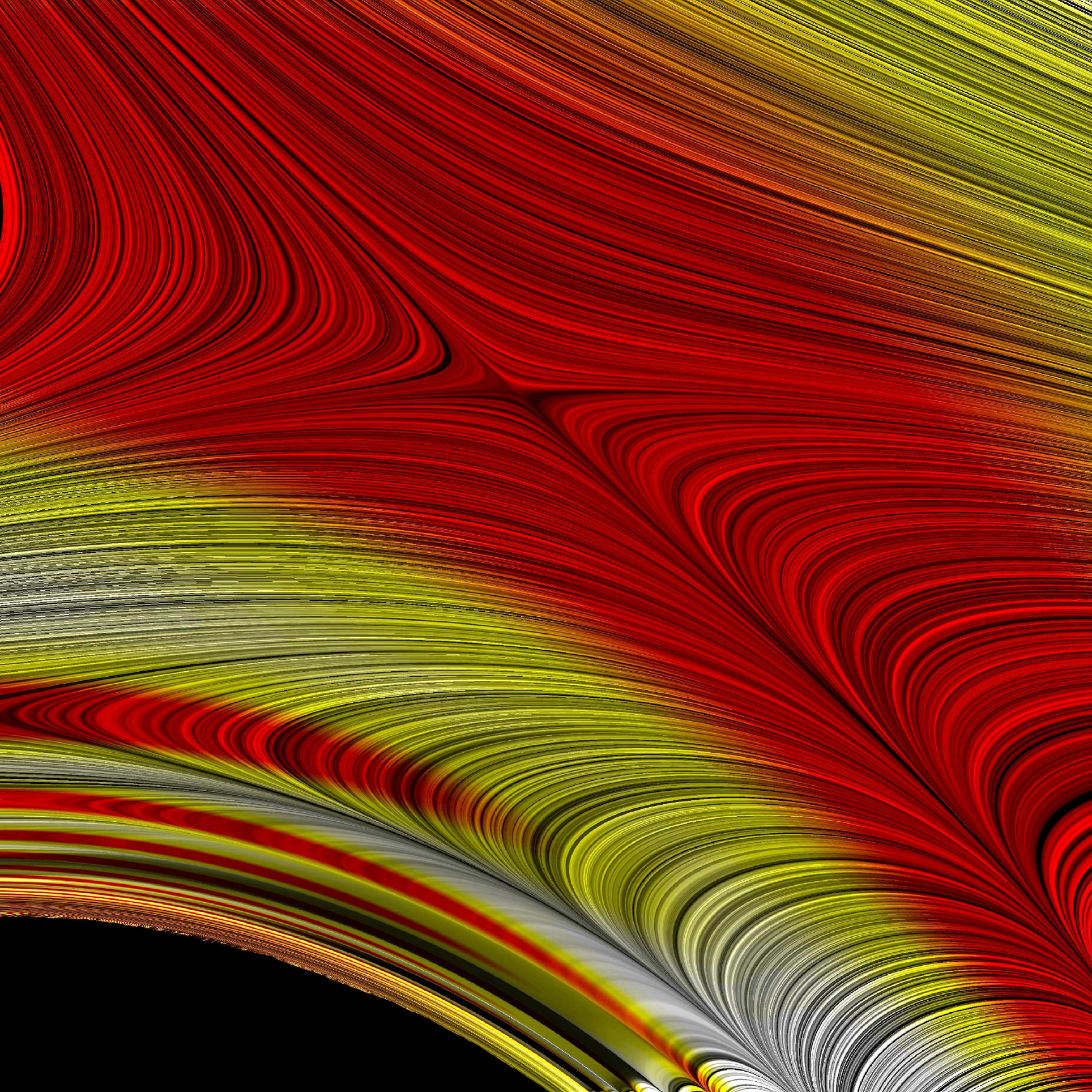}
\caption{The neighborhood of a magnetic null point of 
perpendicular magnetic field in the equatorial plane 
$(x,y)\equiv(r\cos\phi,r\sin\phi)$, i.e., viewed along the rotation 
axis of an extreme ($a=M$) black hole. 
The black circular section in the bottom left corner denotes 
the horizon, $r=r_+(a)$. 
The colour scale indicates the magnetic intensity (in arbitrary 
units).}
\label{fig3}
\end{figure}

\begin{figure}[tbh!]
\vspace*{-0.9em}
\centering
\hfill
\includegraphics[width=0.48\textwidth]{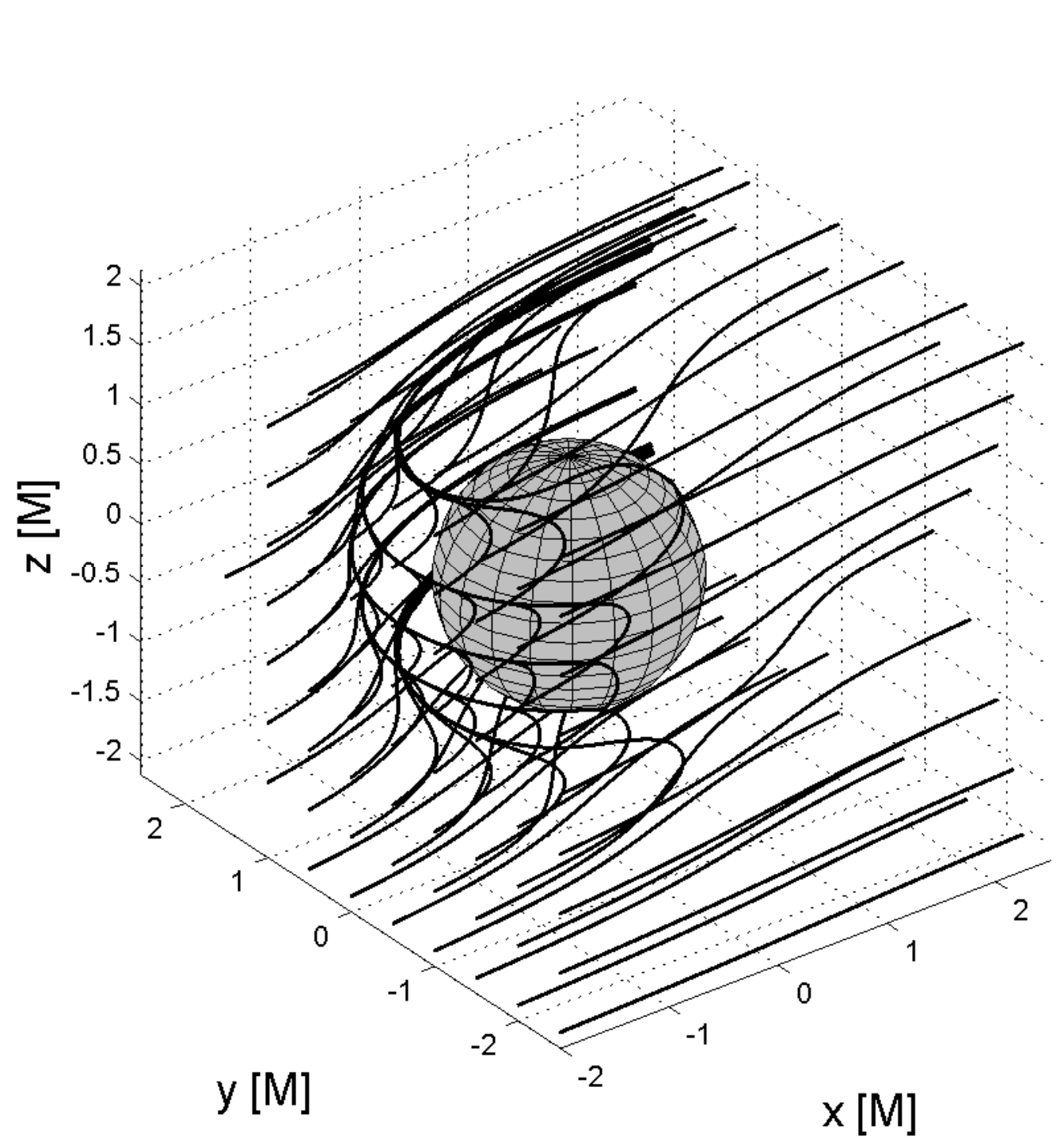}
\hfill
\includegraphics[width=0.44\textwidth]{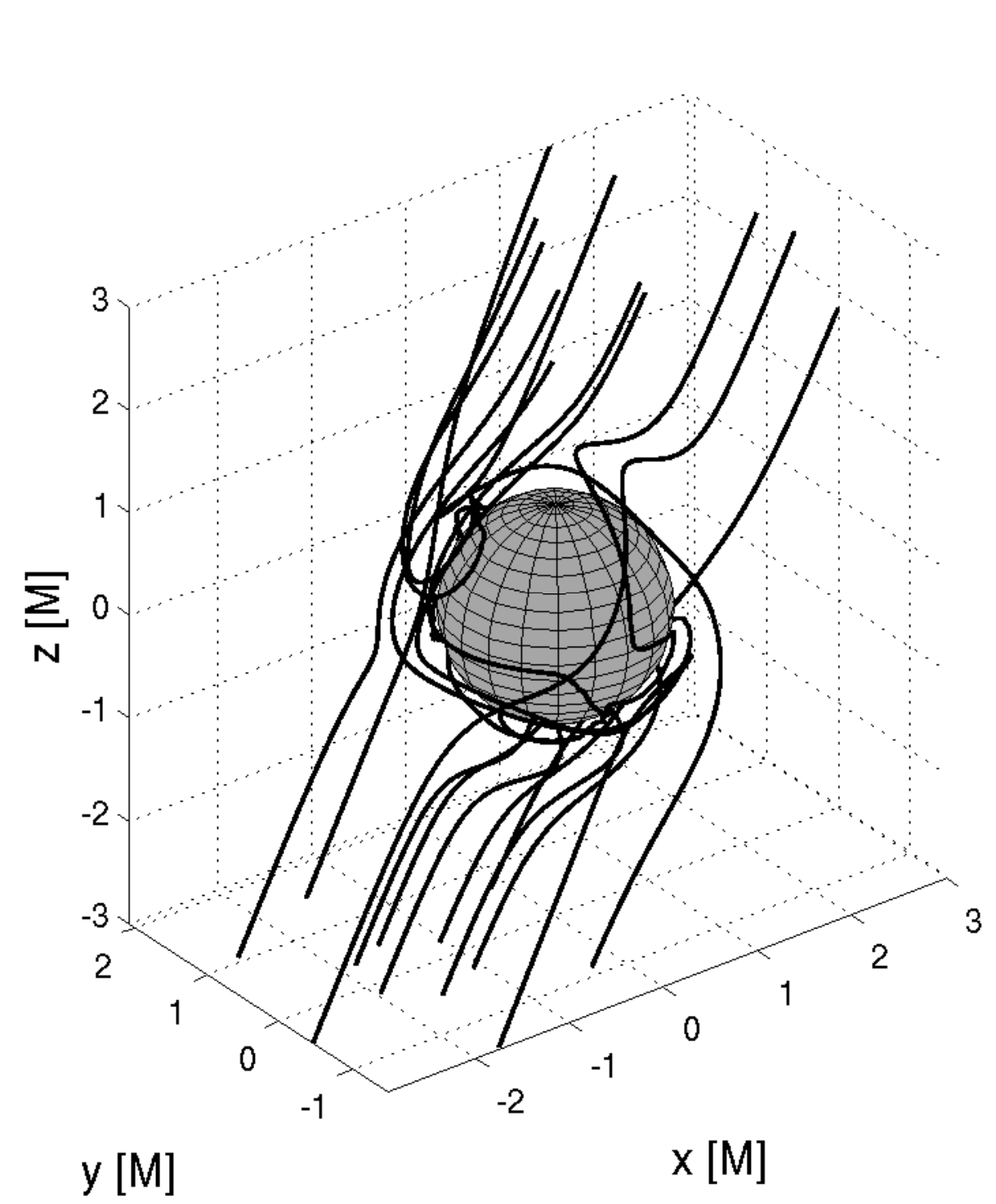}
\hfill~
\caption{Left: Magnetic field asymptotically perpendicular to the 
rotation axis ($z$) of an $a=M$ black hole, centered on the origin. 
Right: Similar example as on the left side, but with the field in a general direction with
respect to rotation axis. }
\label{fig4}
\vspace*{-0em}
\centering
\hfill
\includegraphics[width=0.35\textwidth]{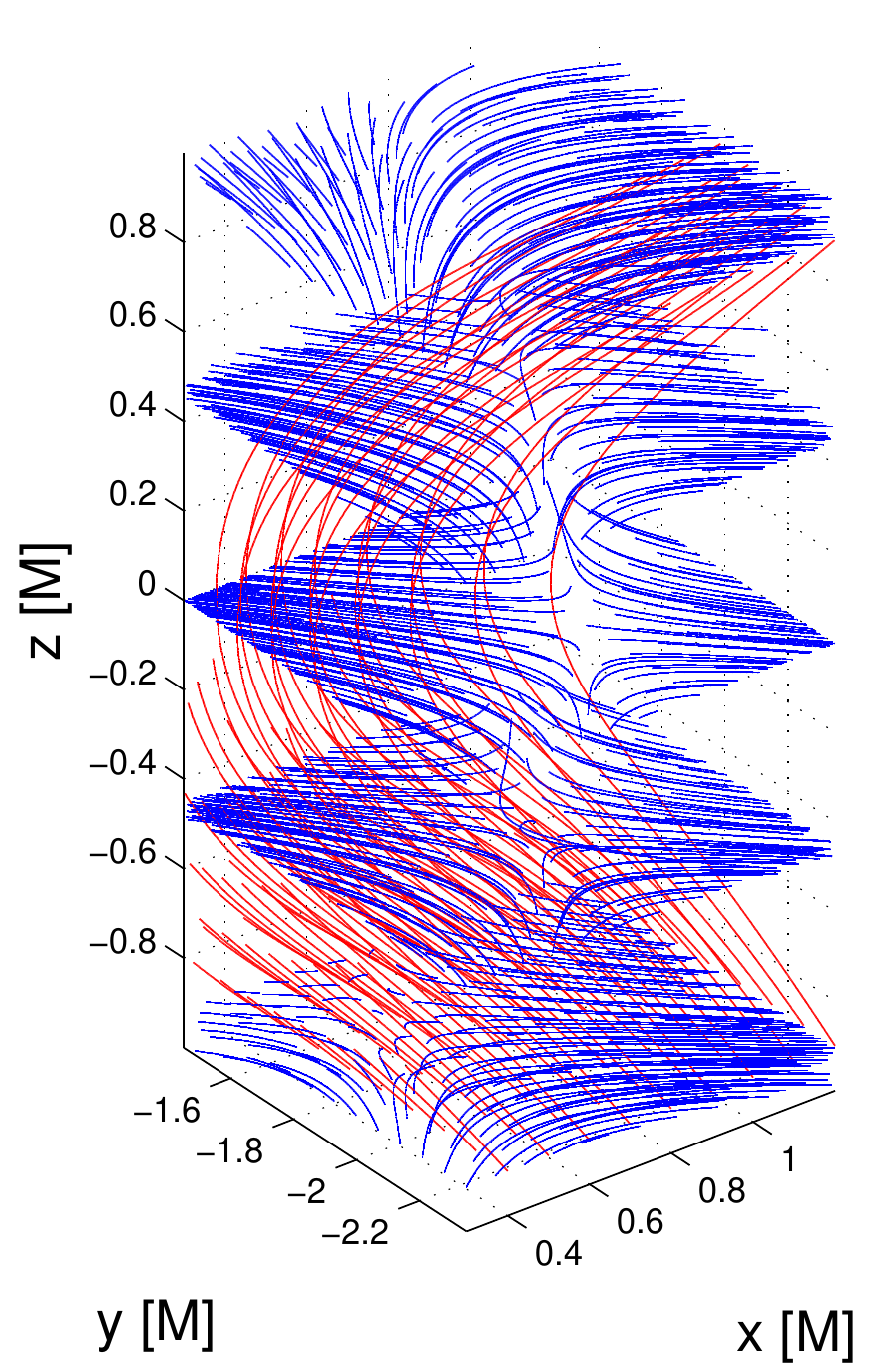}
\hfill~~~~
\includegraphics[width=0.5\textwidth]{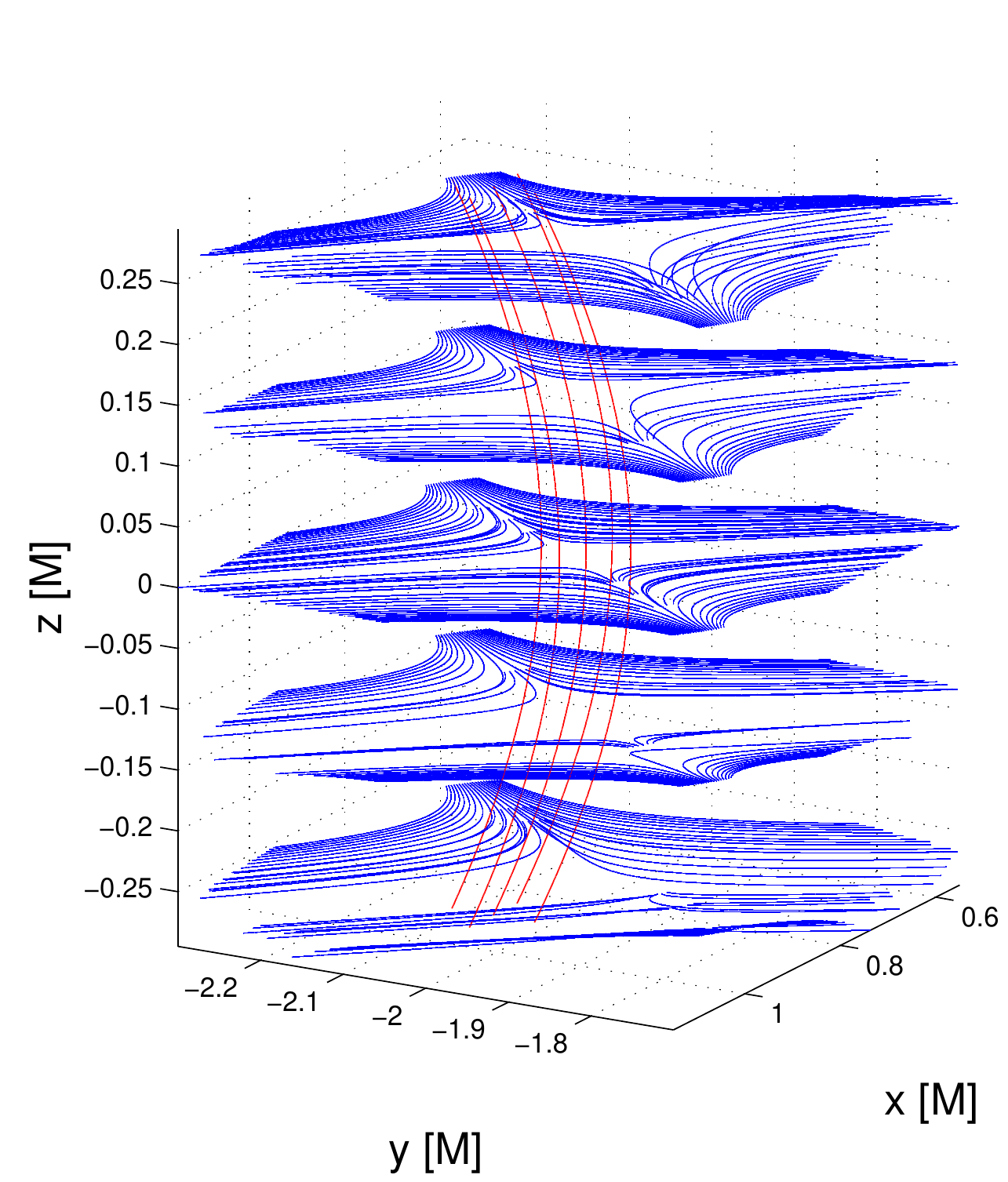}
\hfill~
\caption{Three-dimensional detail around the equatorial plane. 
Magnetic lines are plotted in blue color. The equatorial lines reside within
the plane $z=0$, revealing a null point in the origin. The lines adopt some non-zero
vertical component outside the equatorial plane.
Electric lines are shown in red; they pass through 
the magnetic null point, crossing the equatorial plane in the vertical direction.}
\label{fig5}
\end{figure}

In absence of the perpendicular 
component ($B_{\perp}=0$), the field is relatively uncomplicated 
(see figure \ref{fig1}). Although the frame-dragging acts also on the 
aligned field lines, their shape can be integrated to find the 
surfaces of constant magnetic flux in a closed (analytical) form.
Previously, the aligned fields were explored especially in context 
of magnetic field expulsion from a maximally rotating black hole. 
In our notation, the example in Fig.\ \ref{fig1}
refers to zero velocity of the translatory boost, i.e.\ $\beta=0$. 

\begin{figure}[tbh!]
\centering
\hfill
\includegraphics[width=0.43\textwidth]{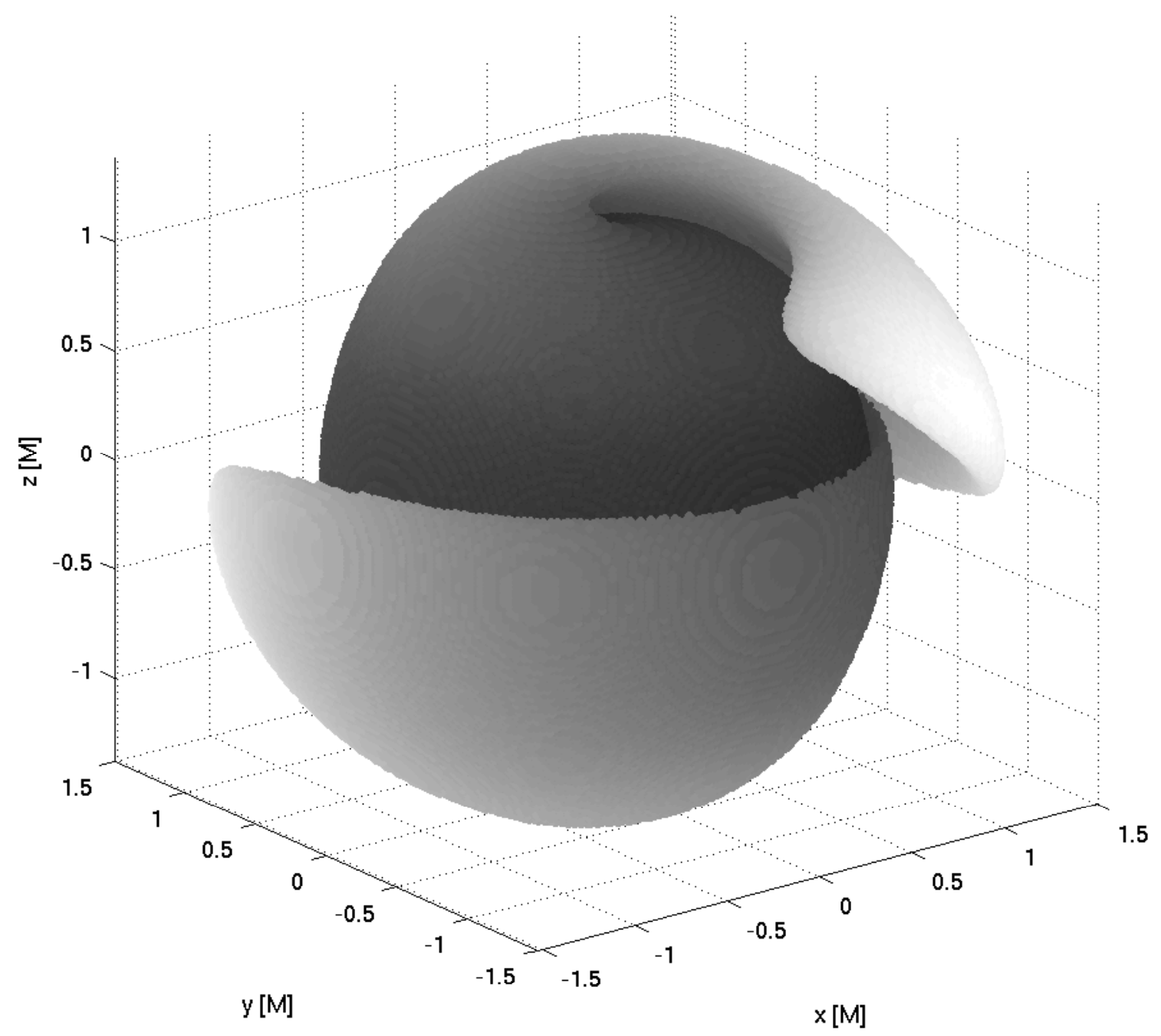}
\hfill
\includegraphics[width=0.43\textwidth]{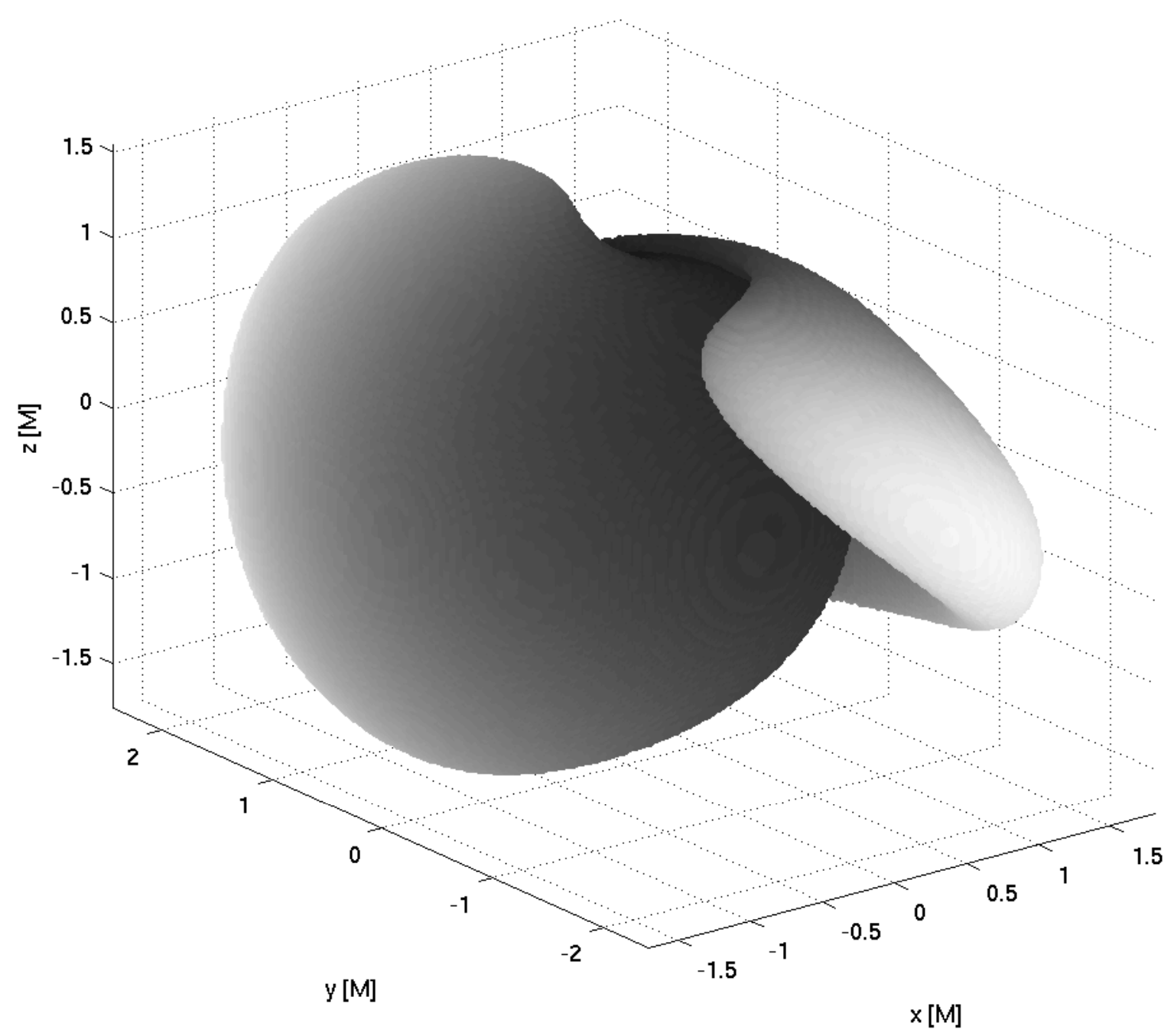}
\hfill~\\
\hfill
\includegraphics[width=0.43\textwidth]{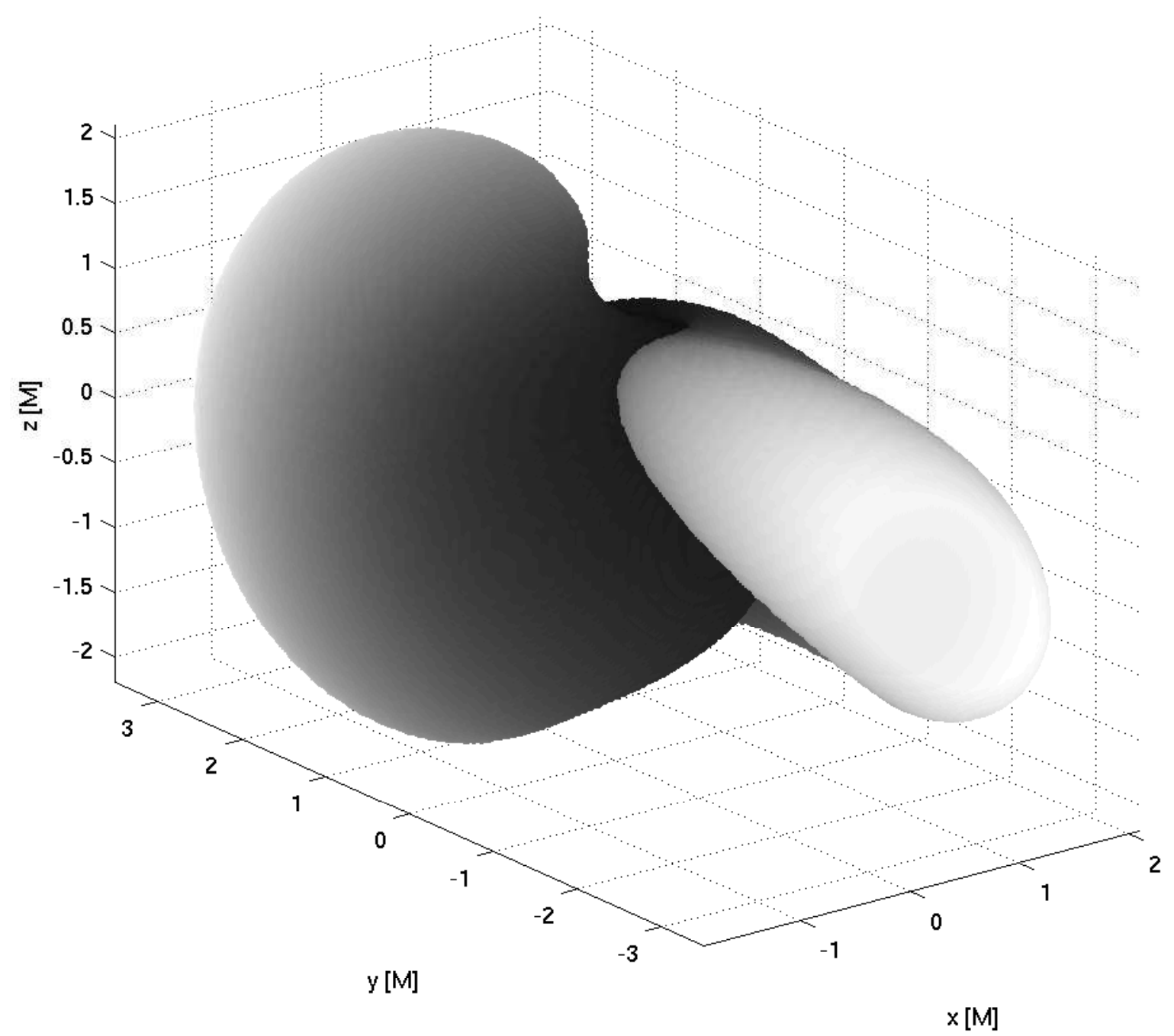}
\hfill
\includegraphics[width=0.43\textwidth]{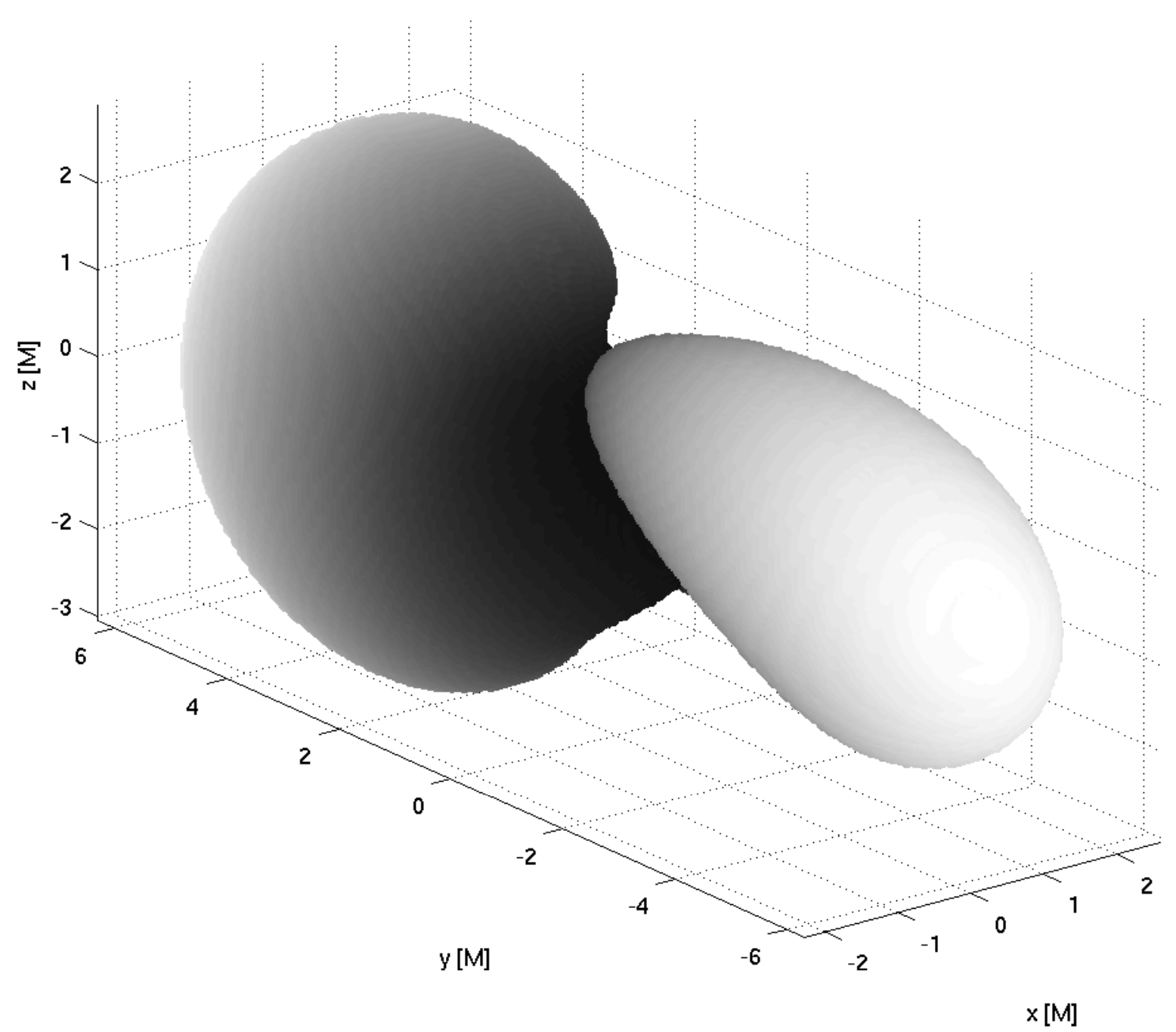}
\hfill~
\caption{Three-dimensional view of the surfaces of equal values of local electric and 
magnetic intensities,
for the case of an oblique background magnetic field $B_x=B_z$ and the extremal 
spin $a=M$ of the rotating Kerr black hole. Boost velocity of the linear motion
is (from upper left to bottom right panels): $v=0$, 0.6, 0.8, and 0.9. Different 
levels of shading of the surface correspond to the distance from origin 
(in Boyer-Lindquist coordinates), where the black hole is located. Increasing
velocity of the translatory motion causes a growing deformation of the 
surface.}
\label{fig6}
\end{figure}

Once a nonzero boost velocity is included, 
the structure of the aligned field becomes more complicated 
(figure \ref{fig2}). This is mainly due to an interplay
between the boost of the black hole and its rotation acting jointly 
on the (originally) aligned field. As $\beta$ increases, the magnetic 
lines are progressively puffed out of horizon and wound around it
(see \citeauthor{k11} \citeyear{k11}, for more examples and details).
\subsection{Neutral points of the magnetic field}
We explore the case of magnetic field with a non-vanishing 
component inclined with respect to rotation axis ($B_{\perp}\neq0$). 
In fact, \citet{kk09} explored a strictly perpendicular case. 
Confining the magnetic lines in the equatorial plane 
$\theta=\pi/2$, the nested structure of magnetic layers 
emerges. These are essential for the magnetic 
reconnection. 

Near-horizon structure of magnetic lines is visualized in 
figure~\ref{fig3} by using the Line-Integral-Convolution 
(LIC) method in Matlab. This technique allows us to identify
clearly the location of neutral points. It turns out to be
particularly useful here with the general orientation of
the asymptotic field direction, as the global solution for the field
lines is too cumbersome. Further,
by introducing $\xi(r;a)\equiv1-r_+(a)/r$ as a new radial coordinate, 
the horizon surface $r=r_+$ collapses into the center and the 
layered structure of magnetic sheets is seen in more detail.

The fully three-dimensional structure of magnetic lines 
develops outside the equatorial plane. In figures~\ref{fig4}--\ref{fig5}
we observe a superposition of two essential effects. Firstly, 
the X-type structure of the magnetic null point persists also
outside the equatorial plane. Secondly,
however, the magnetic lines acquire a growing vertical 
component $B_z$, whereas the 
electric field passes through the magnetic null point and 
crosses the equatorial plane vertically. Such a structure
suggests that particles can be accelerated by the non-vanishing
electric field, and they can stream away from the neutral point.

Finally, another useful way of visualizing the electromagnetic 
structure is by studying the electromagnetic invariants. These can help
us not only to track interesting points of the electromagnetic field
but also to identify locations
of where charges can be electromagnetically supported
or produced via electron-positron pair creation \citep[e.g.,][]{rw75,kv91}.
Recently, \citet{l11} explored the shape of surfaces where the magnitude 
of the local electric field equates the magnetic field component  near a 
magnetized Schwarzschild black hole. 
Figure~\ref{fig6} shows these $E=B$ surfaces in our system.
Obviously the presence of the black hole as well as its motion and
rotation influence the form of these surfaces.

\section{Discussion and Conclusions}
Rotation is a highly interesting attribute of cosmic black 
holes. In principle, black holes can be spun up close 
to extreme $a=M$.
In stellar-mass black holes the spin is thought 
to be chiefly natal \citep{ms06}, whereas supermassive black holes 
in galactic cores can adjust their angular momentum by accretion, and the
outcome of evolution depends largely on the dominant mode of accretion, 
during their entire life-time. In both cases, the spin is an 
important characteristic, potentially allowing the efficient acceleration 
of matter. We showed that it can be also relevant directly for the onset 
of magnetic reconnection.

We discussed the structure of electromagnetic test-fields 
and the layered pattern of current sheets
that can be induced by the gravito-magnetic action. 
Neutral points of the magnetic field 
suggest that magnetic reconnection 
can occur. The proposed scenario can be 
astrophysically relevant in circumstances when the black
hole rotates and moves across a magnetic flux tube, origin of which 
is in currents flowing in the cosmic medium far from the 
black hole. We considered the limit of a magnetically dominated 
system in which the organised magnetic field dictates the 
motion to plasma; the opposite limit of a black hole moving through
a force-free plasma has been investigated by other authors
\citep[recently,][]{pg10,tch11}.

In this paper we considered an idealised situation, starting
from an electro-vacuum solution, assuming fast motion and rotation
of the black hole, and embedding it in an asymptotically uniform
magnetic field. Future simulations should clarify, whether
the astrophysically realistic effects of the moving 
black hole on the surrounding electromagnetic structure in its 
immediate neighborhood will be similar to those envisaged here. 
Despite the field-line structure in this paper being induced
purely by the action of frame-dragging, the exact choice of the 
projection tetrad is not very essential for the existence of 
magnetic layers. The choice of the physical frame
does affect the presence and the exact location of 
the magnetic neutral point, and the associated electric field
which accelerates charged particles away from the
neutral point. Although the exact location of the neutral point
varies with the model parameters, it always occurs very close to 
the black hole, where the frame-dragging is efficient. Therefore,
the point of particle acceleration has to be close
to horizon as well.

\smallskip

We thank the ESA PECS grant No.\ 98040 and the Czech Science
Foundation grant No.\ 205/09/H033 for a continued support.


\end{document}